\documentclass[twocolumn,aps,prx,twoside,superscriptaddress,floatfix,longbibliography,nofootinbib]{revtex4-2}

\usepackage{dcolumn}
\usepackage{float}
\usepackage{bm}
\usepackage[T1]{fontenc}
\usepackage{subcaption}
\usepackage{amsmath}
\usepackage{graphicx} 
\usepackage{tabularx}
\usepackage{multirow}
\usepackage{rotating} 
\usepackage{makecell}
\usepackage{hyperref}
\hypersetup{colorlinks=true, linkcolor=blue, citecolor=blue, urlcolor=blue}
\usepackage{tikz}
\usepackage[normalem]{ulem}

\usepackage[utf8]{inputenc}
\usepackage{booktabs}

\usepackage{color}

\begin{document}
\title{Towards a Modern Theory of Chiralization} 

\author{Nicola A. Spaldin}
\affiliation{Materials Theory, ETH Zurich, Wolfgang-Pauli-Strasse 28, 8093 Z\"urich, Switzerland}

\date{\today}
\begin{abstract}
The Modern Theory of Polarization, which rigorously defines the spontaneous electric polarization of a
periodic solid and provides a recipe for its computation in electronic structure codes, transformed our understanding of ferroelectricity and related dielectric properties. Here we call for the development of an analogous Modern Theory of Chiralization. We review earlier attempts to quantify chirality, highlight the fundamental and practical developments that a modern theory would facilitate, and suggest possible promising routes to its establishment.
\end{abstract}
\maketitle
 
 Chirality underpins an astonishing diversity of domains, ranging from the origin of life via the handedness of DNA, through pharmaceutical drug development using enantiomeric molecules, to $CP$ violation in elementary particles. In terms of applications, chiral optical activity is already widely used in displays and sensors, and recently discovered phenomena such as chiral-induced spin selectivity and exotic chiral topologies hold potential for novel energy-efficient nanoelectronic devices. Although it has long been known how to classify something as chiral or achiral, quantitative micro- and macroscopic measures of structural chirality are lacking, however, and there is no rigorous recipe for determining whether one material is more or less chiral than another.

\begin{figure}[h] \centerline{\includegraphics[width=1.0\linewidth]{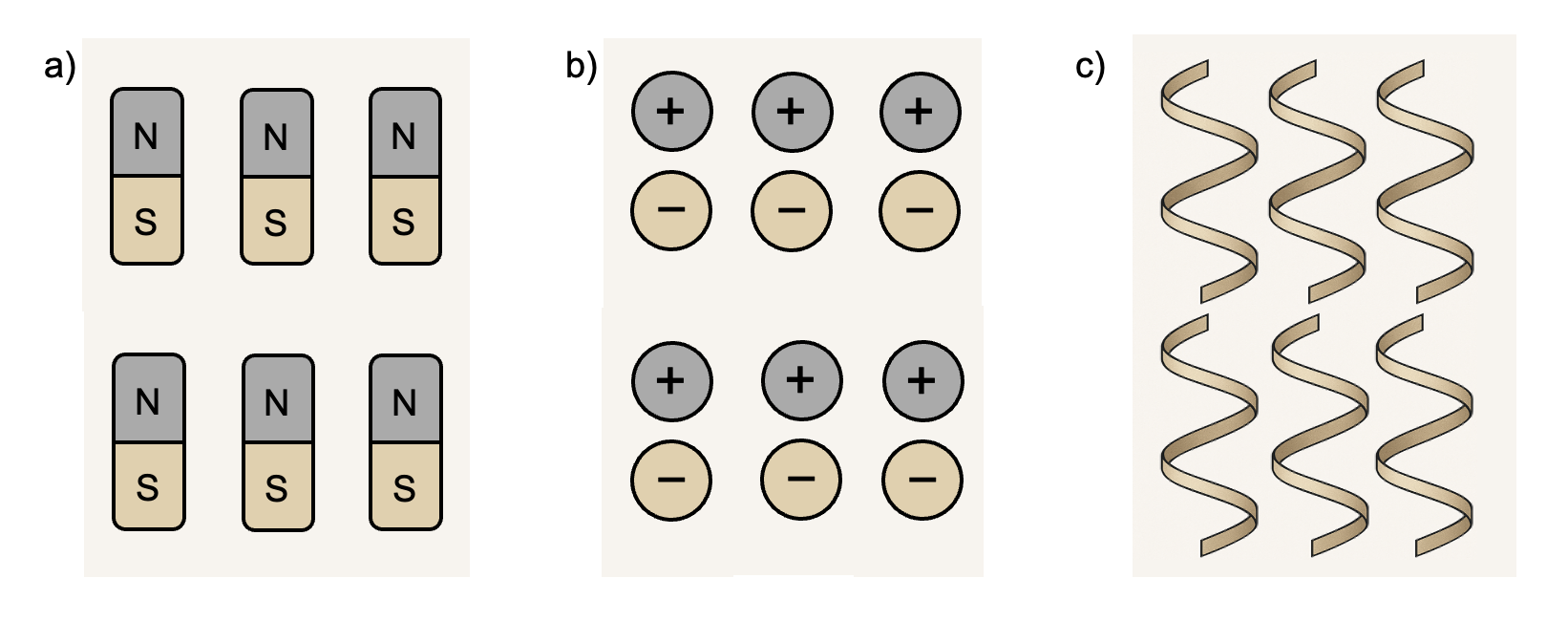}}
    \caption{Cartoon illustrating the analogy between a) magnetic dipole moments and magnetization, b) electric dipole moments and polarization and c) the as-yet unknown local quantity for chirality, which will provide the bulk chiralization.}
    \label{figure1}
\end{figure}

In this commentary, we argue for the development of a formal, quantitative theory of structural chirality, encompassing both a local measure of chirality in finite systems, analogous to the magnetic (electric) dipole moment of magnetic (polar) molecules,  and a bulk thermodynamic chirality per unit volume, or {\it chiralization}, in periodic crystals, analogous to the magnetization (polarization) in bulk ferromagnets (ferroelectrics) (Figure~\ref{figure1}). 

\section*{How would a Modern Theory of Chiralization help us?}

Access to such a chiralization will be particularly valuable in describing so-called ferrochiral materials, which have a symmetry-lowering phase transition from an achiral to a chiral phase. Knowledge of the form of the chiralization, $\chi$, will allow classification of  achiral to ferrochiral phase transitions in terms of an order parameter, and will suggest the development of a Landau theory of structural chirality. By analogy with ferroelectric phase transitions, we might expect the phase transition to be described by a characteristic double well potential energy, as sketched in Figure 2. Here the achiral phase, with $\chi=0$ is at the high-energy peak, and the chiral phases are at the energy minima, with opposite signs of $\chi$ indicating opposite chiral domains. In addition, $\chi$ will determine the conjugate field for chirality, let's call it $\mathbf{Z}$, such that the energy of a chiral system in the presence of its conjugate field is given by 
\begin{equation} \text{Energy} = - \chi \cdot \mathbf{Z}  \quad .
\label{ChiralEnergy}
\end{equation}
Identification of the conjugate field will enable selection of a specific chiral domain, by annealing in the appropriately oriented chiral field that favors the desired orientation of the chiralization.
Finally, knowledge of the form of the chiralization will likely reveal new phenomena associated with chirality, such as unusual properties of surfaces of chiral materials or exotic responses to external perturbations.

\begin{figure}[h] \centerline{\includegraphics[width=1.15\linewidth]{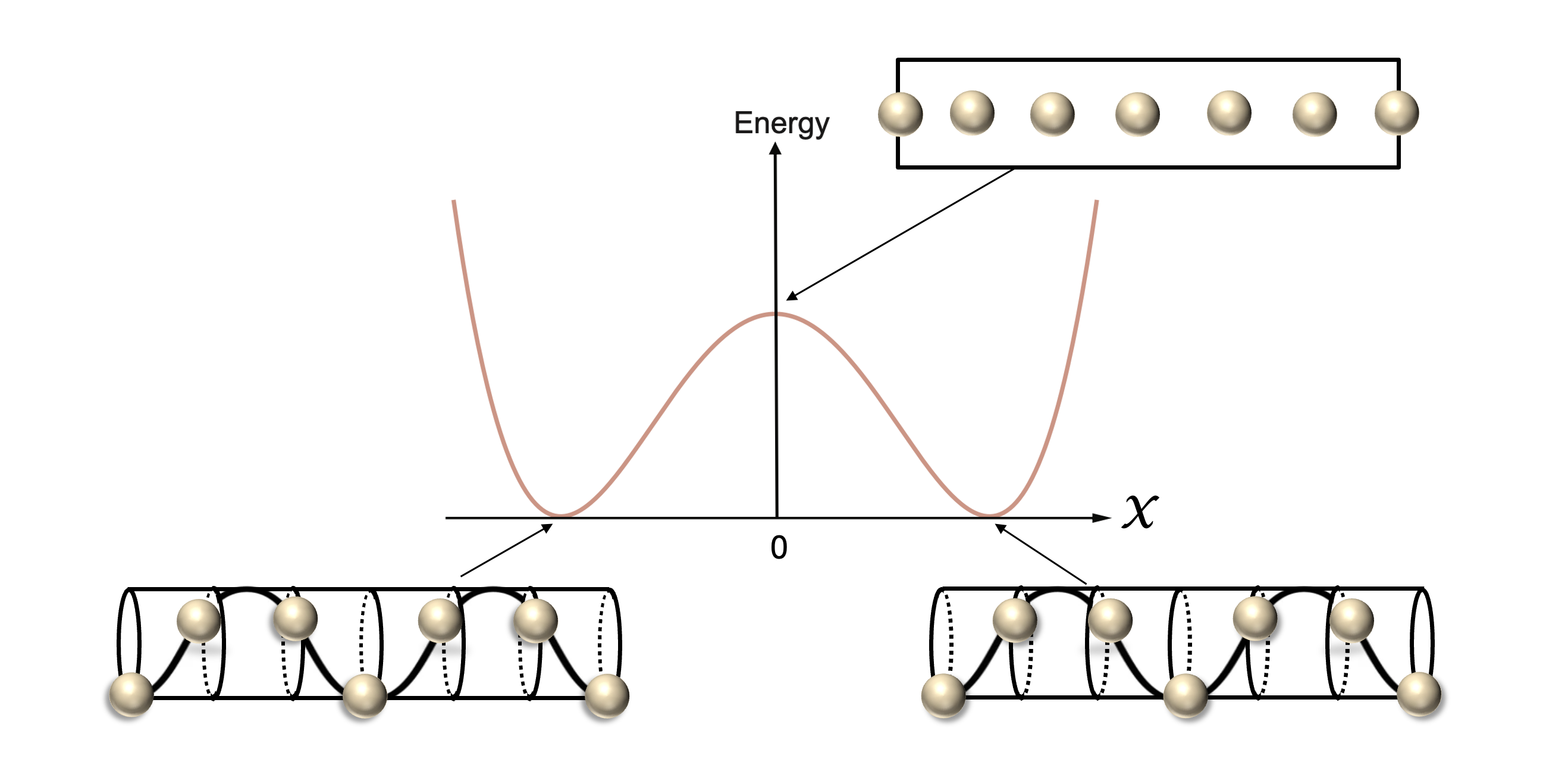}}
    \caption{Cartoon of the anticipated double well potential describing the energy of a ferrochiral phase transition as a function of the chiralization, $\chi$, between the high-temperature, high-symmetry achiral phase ($\chi=0)$ and the low-temperature, low-symmetry chiral structures at the energy minima. }
    \label{figure2}
\end{figure}

\section*{Earlier attempts to quantify chirality}

The question of a rigorous measure of the amount of chirality is a long-standing problem (for reviews see for example Refs.~\cite{Petitjean:2003,Gomez-Ortiz_et_al:2024}). Already in the late 1800s, various {\it asymmetry products}, correlating with the strength of optical rotation and indicating the deviation from a regular tetrahedron, had been proposed \cite{Guye:1893, Brown:1890}. Later, {\it chirality functions}, which facilitate comparisons between chirality magnitudes within a crystal class, were introduced \cite{Ruch/Schoenhofer:1970}, and other measures, based on the similarity between opposite enantiomers \cite{Rassat:1984,Gilat/Schulman:1985} or the distances that the vertices of a shape must move to reach the nearest achiral symmetry point group \cite{Zabrodsky/Avnir:1995}, were discussed. The definition of helicity used in fluid dynamics,  
\begin{equation}
\mathcal{H} = \int d^3\mathbf{r} \, \mathbf{v} \cdot [\nabla \times \mathbf{v}] \quad ,
\end{equation} has recently been proposed \cite{Gomez-Ortiz_et_al:2024} as a measure of structural chirality, with the fluid velocity, $\mathbf{v}$, replaced by the trajectories of the atomic positions between achiral and chiral structures.
This is an intriguing suggestion that could be appropriate for chiral systems that are also helical, but preliminary tests identified known ferrochiral materials for which the sign of $\mathcal{H}$ does not change between enantiomers. In short, none of these measures is completely conceptually satisfactory, with problems such as requiring knowledge of the achiral reference structure and / or yielding the same sign for opposite enantiomers, and they do not provide a formal interpretation of chirality in the sense of Eqn.~\ref{ChiralEnergy}.

\section*{Possible routes towards a modern theory}

\paragraph*{Building on the Modern Theory of Polarizaion.} The chiralization situation today is reminiscent of the state of understanding of the electric polarization before the modern theory of polarization was introduced \cite{Resta:1993,King-Smith/Vanderbilt:1993} more than three decades ago. While the electric dipole of a finite uncharged system is trivially obtained from the expectation value of the position operator $\hat{\mathbf{r}}$, the polarization of a crystal can not be defined as an analogous dipole per unit cell since $\hat{\mathbf{r}}$ is not a well-defined operator in the presence of periodic boundary conditions. This led to a debate about whether polarization, and related properties such as piezoelectricity, are, in fact, bulk or surface properties. The breakthrough involved first realizing that polarization {\it differences} can be obtained from the time-integral of the adiabatic current and later that the polarization itself is not a vector but a lattice of values whose ambiguity is resolved by the surface termination. We are somewhat worse off than our polarization colleagues of the 1990s, however, since we don't even know the chiral analogue to the electric dipole for a finite system.  

\paragraph*{Identifying a chiral multipole.} So, where should we start our quest for a modern theory of chiralization? Well, a first step is to identify a local electromagnetic multipole, analogous to the magnetic or electric dipoles sketched in Fig. 1, that describes a local chiral moment. Here, a useful starting point is the related, simpler class of materials, the space-inversion-symmetric {\it ferroaxials}, for which the electric toroidal dipole, $\mathbf{G}_1$, is established as an order parameter \cite{Bhowal/Spaldin_ETD:2024}. The electric toroidal dipole is an axial vector that is invariant under both time reversal and space inversion. It can be constructed either from a vortex of electric dipole moments, $\mathbf{p}$, as  $\mathbf{G}_1 \sim \sum_i \mathbf{r}_i \times \mathbf{p}_i$, or defined in terms of the orbital and spin angular momenta, $\mathbf{l}$ and $\mathbf{s}$ as $\mathbf{G}_1 \sim \mathbf{l} \times \mathbf{s}$. Ferroaxials are protochiral in the sense that they break all mirror symmetries that contain the electric toroidal dipole but allow mirror symmetry perpendicular to this axis, and tracking the evolution to the chiral phase as this final symmetry is broken might facilitate extension of the formalism to the chiral state.  

Indeed, a leading contemporary proposal for the local multipole describing chirality is the pseudoscalar monopole, $G_0$, of the electric toroidal dipole, $\mathbf{G}_1$. $G_0$ is defined as \cite{Inda_et_al:2024}
\begin{equation} G_0 = \int_V \nabla \cdot \mathbf{G}_1  \space d^3\mathbf{r}
\end{equation} and transforms as required for a chiral quantity under time-reversal (even) and space-inversion (odd) symmetries. However, it suffers, like all pseudoscalar quantities, from so-called false zeros: Since a suitable pseudoscalar descriptor must have opposite sign in the two enantiomers, it has to pass through zero somewhere along the pathway transforming one enantiomer into the other, even if the structure remains chiral along the entire path. A possible solution could be found in a multi-component descriptor, for example in the higher order multipoles of the electric toroidal dipole, such as the quadrupole $G_2$, or hexadecapole, $G_4$, or in composites of conventional charge multipoles \cite{Winkler/Zuelicke:2024}. Note that the appropriate local multipole is likely to include a factor of position in its definition, so the extension to a bulk crystal chiralization will pose similar challenges to those encountered in the modern theory of polarization described above \cite{King-Smith/Vanderbilt:1993,Resta:1993}.

\paragraph*{Chiral phonons.}Another possible space to look for insights into the construction of a macroscopic chiral order parameter is in the rapidly advancing field of chiral phonons, which are quantized lattice vibrations that carry angular momentum and break improper rotational symmetry \cite{Juraschek_et_al:2025}. An obvious direction is to search computationally for unstable chiral phonons in achiral structures. Unstable (imaginary frequency) phonons indicate an absence of restoring forces to atomic displacements, and are a feature of structures at the high energy peaks of double well potentials (Figure 2). Therefore, they will likely point to the existence of stable chiral phases, in the same way that polar phonon instabilities in the high symmetry structures of ferroelectrics indicate the eigenvectors of atomic displacements that produce low symmetry polar structures. In
ferroelectrics, the pattern of atomic displacements, $\delta d_j$, connecting a polar structure to its high-symmetry non-polar reference phase, maps directly onto the rigorous definition of polarization given by the modern theory of polarization: 
\begin{equation}
\partial P_i = \frac{e}{\Omega} Z^*_{ij} \delta d_j \quad .
\end{equation} 
Here $e$ is the electronic charge, $\Omega$ the unit cell volume, $Z^*_{ij}$ the Born effective charge tensor on each atom, which can be calculated using the modern theory of polarization or measured from the splitting between longitudinal and transverse optical phonons, and $\partial P_i$ the change in polarization from the centrosymmetric phase, with $i,j$ indicating cartesian or lattice directions. Identification of an analogous ``effective charge'' for the case of chiral systems would lead directly to a macroscopic chiralization. Importantly, phonon chirality, $\bm{S}$, has an established definition \cite{Zhang/Niu:2015}, with, for example, the $z$ component given by
\begin{equation}
    S_z = \sum_{m=1}^n 
    \left(  
    |\langle r_{m,z} | \epsilon_m \rangle |^2 
    - |\langle l_{m,z} | \epsilon_m \rangle |^2 
    \right) \quad .
    \label{eqn:phonon_chirality}
\end{equation} 
Here, $\epsilon$ is the normalized phonon eigenvector, $r$ and $l$ are eigenvectors corresponding to pure right- and left-handed rotations, and the sum is over all the atoms of the unit cell. By extending this definition to atomic displacements rather than vibrations (using normal rather than phonon modes), the mode chirality of Eqn.~\ref{eqn:phonon_chirality} could be modified to provide an order parameter for primary ferrochiral transitions. 

We hope that the considerations outlined here motivate and facilitate progress towards a modern theory of chiralization, inspire the development of experimental tools that will enable its verification, and encourage exploration of novel phenomena that the theory might indicate. While our focus here is on structural chirality, a successful theory might also encompass more exotic forms such as chiral spin systems or topological chirality and reveal as yet unseen connections between them. 

\section*{Acknowledgements} NS acknowledges financial support from the Swiss National Science Foundation, project number TMAG-2\_225790 {\it Static and Dynamic Crystal Chirality}.


\begin{thebibliography}{15}%
\makeatletter
\providecommand \@ifxundefined [1]{%
 \@ifx{#1\undefined}
}%
\providecommand \@ifnum [1]{%
 \ifnum #1\expandafter \@firstoftwo
 \else \expandafter \@secondoftwo
 \fi
}%
\providecommand \@ifx [1]{%
 \ifx #1\expandafter \@firstoftwo
 \else \expandafter \@secondoftwo
 \fi
}%
\providecommand \natexlab [1]{#1}%
\providecommand \enquote  [1]{``#1''}%
\providecommand \bibnamefont  [1]{#1}%
\providecommand \bibfnamefont [1]{#1}%
\providecommand \citenamefont [1]{#1}%
\providecommand \href@noop [0]{\@secondoftwo}%
\providecommand \href [0]{\begingroup \@sanitize@url \@href}%
\providecommand \@href[1]{\@@startlink{#1}\@@href}%
\providecommand \@@href[1]{\endgroup#1\@@endlink}%
\providecommand \@sanitize@url [0]{\catcode `\\12\catcode `\$12\catcode `\&12\catcode `\#12\catcode `\^12\catcode `\_12\catcode `\%12\relax}%
\providecommand \@@startlink[1]{}%
\providecommand \@@endlink[0]{}%
\providecommand \url  [0]{\begingroup\@sanitize@url \@url }%
\providecommand \@url [1]{\endgroup\@href {#1}{\urlprefix }}%
\providecommand \urlprefix  [0]{URL }%
\providecommand \Eprint [0]{\href }%
\providecommand \doibase [0]{https://doi.org/}%
\providecommand \selectlanguage [0]{\@gobble}%
\providecommand \bibinfo  [0]{\@secondoftwo}%
\providecommand \bibfield  [0]{\@secondoftwo}%
\providecommand \translation [1]{[#1]}%
\providecommand \BibitemOpen [0]{}%
\providecommand \bibitemStop [0]{}%
\providecommand \bibitemNoStop [0]{.\EOS\space}%
\providecommand \EOS [0]{\spacefactor3000\relax}%
\providecommand \BibitemShut  [1]{\csname bibitem#1\endcsname}%
\let\auto@bib@innerbib\@empty
\bibitem [{\citenamefont {Petitjean}(2003)}]{Petitjean:2003}%
  \BibitemOpen
  \bibfield  {author} {\bibinfo {author} {\bibfnamefont {M.}~\bibnamefont {Petitjean}},\ }\bibfield  {title} {\bibinfo {title} {Chirality and symmetry measures: A transdisciplinary review},\ }\href@noop {} {\bibfield  {journal} {\bibinfo  {journal} {Entropy}\ }\textbf {\bibinfo {volume} {5}},\ \bibinfo {pages} {271} (\bibinfo {year} {2003})}\BibitemShut {NoStop}%
\bibitem [{\citenamefont {G{\'o}mez-Ortiz}\ \emph {et~al.}(2024)\citenamefont {G{\'o}mez-Ortiz}, \citenamefont {Fava}, \citenamefont {McCabe}, \citenamefont {Romero},\ and\ \citenamefont {Bousquet}}]{Gomez-Ortiz_et_al:2024}%
  \BibitemOpen
  \bibfield  {author} {\bibinfo {author} {\bibfnamefont {F.}~\bibnamefont {G{\'o}mez-Ortiz}}, \bibinfo {author} {\bibfnamefont {M.}~\bibnamefont {Fava}}, \bibinfo {author} {\bibfnamefont {E.~E.}\ \bibnamefont {McCabe}}, \bibinfo {author} {\bibfnamefont {A.~H.}\ \bibnamefont {Romero}},\ and\ \bibinfo {author} {\bibfnamefont {E.}~\bibnamefont {Bousquet}},\ }\bibfield  {title} {\bibinfo {title} {Pros and cons of structural chirality measurements and computation of handedness in periodic solids},\ }\bibfield  {journal} {\bibinfo  {journal} {arXiv preprint 2405.16268}\ }\href {https://doi.org/10.48550/arXiv.2405.16268} {10.48550/arXiv.2405.16268} (\bibinfo {year} {2024})\BibitemShut {NoStop}%
\bibitem [{\citenamefont {Guye}(1893)}]{Guye:1893}%
  \BibitemOpen
  \bibfield  {author} {\bibinfo {author} {\bibfnamefont {P.-A.}\ \bibnamefont {Guye}},\ }\bibfield  {title} {\bibinfo {title} {Sur le produit d'asym\'etrie},\ }\href@noop {} {\bibfield  {journal} {\bibinfo  {journal} {Compt. Rend. Hebdom. Acad. Sci.}\ }\textbf {\bibinfo {volume} {16}},\ \bibinfo {pages} {1378} (\bibinfo {year} {1893})}\BibitemShut {NoStop}%
\bibitem [{\citenamefont {Brown}(1890)}]{Brown:1890}%
  \BibitemOpen
  \bibfield  {author} {\bibinfo {author} {\bibfnamefont {C.}~\bibnamefont {Brown}},\ }\bibfield  {title} {\bibinfo {title} {On the relation of optical activity to the character of the radicals united to the asymmetric carbon atom},\ }\href@noop {} {\bibfield  {journal} {\bibinfo  {journal} {Proc. Roy. Soc. Edinburgh}\ }\textbf {\bibinfo {volume} {17}},\ \bibinfo {pages} {181} (\bibinfo {year} {1890})}\BibitemShut {NoStop}%
\bibitem [{\citenamefont {Ruch}\ and\ \citenamefont {Schoenhofer}(1970)}]{Ruch/Schoenhofer:1970}%
  \BibitemOpen
  \bibfield  {author} {\bibinfo {author} {\bibfnamefont {E.}~\bibnamefont {Ruch}}\ and\ \bibinfo {author} {\bibfnamefont {A.}~\bibnamefont {Schoenhofer}},\ }\bibfield  {title} {\bibinfo {title} {{Theorie der Chiralit\"atsfunktionen}},\ }\href@noop {} {\bibfield  {journal} {\bibinfo  {journal} {Theor. Chim. Acta}\ }\textbf {\bibinfo {volume} {19}},\ \bibinfo {pages} {225} (\bibinfo {year} {1970})}\BibitemShut {NoStop}%
\bibitem [{\citenamefont {Rassat}(1984)}]{Rassat:1984}%
  \BibitemOpen
  \bibfield  {author} {\bibinfo {author} {\bibfnamefont {A.}~\bibnamefont {Rassat}},\ }\bibfield  {title} {\bibinfo {title} {{Un crit\'ere de classement des syst\`emes chiraux de points \`a partir de la distance au sens de Haussdorf}},\ }\href@noop {} {\bibfield  {journal} {\bibinfo  {journal} {Compt. Rend. Acad. Sci. Paris (Serie II)}\ }\textbf {\bibinfo {volume} {299}},\ \bibinfo {pages} {53} (\bibinfo {year} {1984})}\BibitemShut {NoStop}%
\bibitem [{\citenamefont {Gilat}\ and\ \citenamefont {Schulman}(1985)}]{Gilat/Schulman:1985}%
  \BibitemOpen
  \bibfield  {author} {\bibinfo {author} {\bibfnamefont {G.}~\bibnamefont {Gilat}}\ and\ \bibinfo {author} {\bibfnamefont {L.~S.}\ \bibnamefont {Schulman}},\ }\bibfield  {title} {\bibinfo {title} {{Chiral interaction, magnitude of the effects and application to natural selection of L-enantiomer}},\ }\href@noop {} {\bibfield  {journal} {\bibinfo  {journal} {Chem. Phys. Lett.}\ }\textbf {\bibinfo {volume} {121}},\ \bibinfo {pages} {13} (\bibinfo {year} {1985})}\BibitemShut {NoStop}%
\bibitem [{\citenamefont {Zabrodsky}\ and\ \citenamefont {Avnir}(1995)}]{Zabrodsky/Avnir:1995}%
  \BibitemOpen
  \bibfield  {author} {\bibinfo {author} {\bibfnamefont {H.}~\bibnamefont {Zabrodsky}}\ and\ \bibinfo {author} {\bibfnamefont {D.}~\bibnamefont {Avnir}},\ }\bibfield  {title} {\bibinfo {title} {{Continuous symmetry measures. 4. Chirality}},\ }\href@noop {} {\bibfield  {journal} {\bibinfo  {journal} {J. Am. Chem. Soc.}\ }\textbf {\bibinfo {volume} {117}},\ \bibinfo {pages} {462 } (\bibinfo {year} {1995})}\BibitemShut {NoStop}%
\bibitem [{\citenamefont {Resta}(1993)}]{Resta:1993}%
  \BibitemOpen
  \bibfield  {author} {\bibinfo {author} {\bibfnamefont {R.}~\bibnamefont {Resta}},\ }\bibfield  {title} {\bibinfo {title} {Macroscopic electric polarization as a geometric quantum phase},\ }\href@noop {} {\bibfield  {journal} {\bibinfo  {journal} {Eur. Phys. Lett.}\ }\textbf {\bibinfo {volume} {22}},\ \bibinfo {pages} {133} (\bibinfo {year} {1993})}\BibitemShut {NoStop}%
\bibitem [{\citenamefont {King-Smith}\ and\ \citenamefont {Vanderbilt}(1993)}]{King-Smith/Vanderbilt:1993}%
  \BibitemOpen
  \bibfield  {author} {\bibinfo {author} {\bibfnamefont {R.~D.}\ \bibnamefont {King-Smith}}\ and\ \bibinfo {author} {\bibfnamefont {D.}~\bibnamefont {Vanderbilt}},\ }\bibfield  {title} {\bibinfo {title} {Theory of polarization of crystalline solids},\ }\href@noop {} {\bibfield  {journal} {\bibinfo  {journal} {Phys. Rev. B}\ }\textbf {\bibinfo {volume} {47}},\ \bibinfo {pages} {R1651} (\bibinfo {year} {1993})}\BibitemShut {NoStop}%
\bibitem [{\citenamefont {Bhowal}\ and\ \citenamefont {Spaldin}(2024)}]{Bhowal/Spaldin_ETD:2024}%
  \BibitemOpen
  \bibfield  {author} {\bibinfo {author} {\bibfnamefont {S.}~\bibnamefont {Bhowal}}\ and\ \bibinfo {author} {\bibfnamefont {N.~A.}\ \bibnamefont {Spaldin}},\ }\bibfield  {title} {\bibinfo {title} {Electric toroidal dipole order and hidden spin polarization in ferroaxial materials},\ }\href@noop {} {\bibfield  {journal} {\bibinfo  {journal} {Phys. Rev. Research}\ }\textbf {\bibinfo {volume} {6}},\ \bibinfo {pages} {043141} (\bibinfo {year} {2024})}\BibitemShut {NoStop}%
\bibitem [{\citenamefont {Inda}\ \emph {et~al.}(2024)\citenamefont {Inda}, \citenamefont {Oiwa}, \citenamefont {Hayami}, \citenamefont {Yamamoto},\ and\ \citenamefont {Kusunose}}]{Inda_et_al:2024}%
  \BibitemOpen
  \bibfield  {author} {\bibinfo {author} {\bibfnamefont {A.}~\bibnamefont {Inda}}, \bibinfo {author} {\bibfnamefont {R.}~\bibnamefont {Oiwa}}, \bibinfo {author} {\bibfnamefont {S.}~\bibnamefont {Hayami}}, \bibinfo {author} {\bibfnamefont {H.~M.}\ \bibnamefont {Yamamoto}},\ and\ \bibinfo {author} {\bibfnamefont {H.}~\bibnamefont {Kusunose}},\ }\bibfield  {title} {\bibinfo {title} {Quantification of chirality based on electric toroidal monopole},\ }\href {https://doi.org/10.1063/5.0204254} {\bibfield  {journal} {\bibinfo  {journal} {J. Chem. Phys.}\ }\textbf {\bibinfo {volume} {160}},\ \bibinfo {pages} {184117} (\bibinfo {year} {2024})}\BibitemShut {NoStop}%
\bibitem [{\citenamefont {Winkler}\ and\ \citenamefont {Zuelicke}(2024)}]{Winkler/Zuelicke:2024}%
  \BibitemOpen
  \bibfield  {author} {\bibinfo {author} {\bibfnamefont {R.}~\bibnamefont {Winkler}}\ and\ \bibinfo {author} {\bibfnamefont {U.}~\bibnamefont {Zuelicke}},\ }\bibfield  {title} {\bibinfo {title} {{Standard model of electromagnetism and chirality in crystals}},\ }\href@noop {} {\bibfield  {journal} {\bibinfo  {journal} {arXiv}\ } (\bibinfo {year} {2024})},\ \Eprint {https://arxiv.org/abs/2405.20940} {2405.20940} \BibitemShut {NoStop}%
\bibitem [{\citenamefont {Juraschek}\ \emph {et~al.}(2025)\citenamefont {Juraschek}, \citenamefont {Geilhufe}, \citenamefont {Zhu}, \citenamefont {Basini}, \citenamefont {Baum}, \citenamefont {Baydin}, \citenamefont {Chaudhary}, \citenamefont {Fechner}, \citenamefont {Flebus}, \citenamefont {Grissonnanche}, \citenamefont {Kirilyuk}, \citenamefont {Lemeshko}, \citenamefont {Maehrlein}, \citenamefont {Mignolet}, \citenamefont {Murakami}, \citenamefont {Niu}, \citenamefont {Nowak}, \citenamefont {Romao}, \citenamefont {Rostami}, \citenamefont {Satoh}, \citenamefont {Spaldin}, \citenamefont {Ueda},\ and\ \citenamefont {Zhang}}]{Juraschek_et_al:2025}%
  \BibitemOpen
  \bibfield  {author} {\bibinfo {author} {\bibfnamefont {D.~M.}\ \bibnamefont {Juraschek}}, \bibinfo {author} {\bibfnamefont {R.~M.}\ \bibnamefont {Geilhufe}}, \bibinfo {author} {\bibfnamefont {H.}~\bibnamefont {Zhu}}, \bibinfo {author} {\bibfnamefont {M.}~\bibnamefont {Basini}}, \bibinfo {author} {\bibfnamefont {P.}~\bibnamefont {Baum}}, \bibinfo {author} {\bibfnamefont {A.}~\bibnamefont {Baydin}}, \bibinfo {author} {\bibfnamefont {S.}~\bibnamefont {Chaudhary}}, \bibinfo {author} {\bibfnamefont {M.}~\bibnamefont {Fechner}}, \bibinfo {author} {\bibfnamefont {B.}~\bibnamefont {Flebus}}, \bibinfo {author} {\bibfnamefont {G.}~\bibnamefont {Grissonnanche}}, \bibinfo {author} {\bibfnamefont {A.~I.}\ \bibnamefont {Kirilyuk}}, \bibinfo {author} {\bibfnamefont {M.}~\bibnamefont {Lemeshko}}, \bibinfo {author} {\bibfnamefont {S.~F.}\ \bibnamefont {Maehrlein}}, \bibinfo {author} {\bibfnamefont {M.}~\bibnamefont {Mignolet}}, \bibinfo {author} {\bibfnamefont {S.}~\bibnamefont {Murakami}}, \bibinfo {author} {\bibfnamefont
  {Q.}~\bibnamefont {Niu}}, \bibinfo {author} {\bibfnamefont {U.}~\bibnamefont {Nowak}}, \bibinfo {author} {\bibfnamefont {C.~P.}\ \bibnamefont {Romao}}, \bibinfo {author} {\bibfnamefont {H.}~\bibnamefont {Rostami}}, \bibinfo {author} {\bibfnamefont {T.}~\bibnamefont {Satoh}}, \bibinfo {author} {\bibfnamefont {N.~A.}\ \bibnamefont {Spaldin}}, \bibinfo {author} {\bibfnamefont {H.}~\bibnamefont {Ueda}},\ and\ \bibinfo {author} {\bibfnamefont {L.}~\bibnamefont {Zhang}},\ }\bibfield  {title} {\bibinfo {title} {{Chiral phonons}},\ }\href {https://doi.org/10.1038/s41567-025-03001-9} {\bibfield  {journal} {\bibinfo  {journal} {Nat. Phys.}\ }\textbf {\bibinfo {volume} {21}},\ \bibinfo {pages} {1532} (\bibinfo {year} {2025})}\BibitemShut {NoStop}%
\bibitem [{\citenamefont {Zhang}\ and\ \citenamefont {Niu}(2015)}]{Zhang/Niu:2015}%
  \BibitemOpen
  \bibfield  {author} {\bibinfo {author} {\bibfnamefont {L.}~\bibnamefont {Zhang}}\ and\ \bibinfo {author} {\bibfnamefont {Q.}~\bibnamefont {Niu}},\ }\bibfield  {title} {\bibinfo {title} {Chiral phonons at high-symmetry points in monolayer hexagonal lattices},\ }\href@noop {} {\bibfield  {journal} {\bibinfo  {journal} {Phys. Rev. Lett.}\ }\textbf {\bibinfo {volume} {115}},\ \bibinfo {pages} {115502} (\bibinfo {year} {2015})}\BibitemShut {NoStop}%
\end{thebibliography}
\end{document}